\documentclass[notitlepage,aps,epsfig,showpacs,floats,twocolumn,amssymb,amsmath,floatfix,
groupedaddress,superscriptaddress]{revtex4-1}
\usepackage{graphicx} 
\usepackage{dcolumn} 
\usepackage{bm} 
\usepackage{float} 
\usepackage{bbm} 
\usepackage{enumitem,xcolor}
\usepackage{color}
\usepackage{amsfonts}
\usepackage{dsfont}
\usepackage{xcolor}
\usepackage{lmodern} 
\usepackage{natbib}
\usepackage{comment}
\usepackage{amsmath} 
\usepackage{amsmath,amssymb}
\usepackage{natbib}
\usepackage[margin=2cm]{geometry}
\usepackage{tikz}
\usepackage{ulem}
\usetikzlibrary{matrix}

\newcommand\bi{\begin{itemize}}
\newcommand\ei{\end{itemize}}

\usepackage[colorlinks=true,citecolor=blue]{hyperref}
\hypersetup{colorlinks=true,citecolor=blue,linkcolor=blue,urlcolor=black}

\usepackage{geometry}

\usepackage{tikz} 
\usetikzlibrary{shapes,arrows,calc}
\usetikzlibrary{datavisualization}
\usetikzlibrary{datavisualization.formats.functions}
\usepackage{pgfplots} 
\usepackage{tikz}
\usetikzlibrary{matrix}  

\usepackage[colorlinks=true,citecolor=blue]{hyperref}

\providecommand{\abs}[1]{\lvert#1\rvert}

\newcommand{\dd}[0]{\mathrm{d}}
\newcommand{\tr}[0]{\mathrm{tr}}

\newcommand{\id}[0]{\mathds{1}}
\newcommand{\matr}[1]{\bm{#1}}
\newcommand{\vect}[1]{\vec{#1}}
\newcommand{\uvect}[1]{\hat{#1}}
\newcommand{\transp}{\mathsf{T}}
\newcommand{\vecnabla}{\vect\nabla}

\begin{document}
\title{Phase field models of active matter}
\author{Romain Mueller}
\affiliation{Rudolf Peierls Centre for Theoretical Physics, University of Oxford, Parks Road, OX1 3NP, Oxford, United Kingdom}
\author{Amin Doostmohammadi}
\email{doostmohammadi@nbi.ku.dk}
\affiliation{The Niels Bohr Institute, University of Copenhagen, Copenhagen, Denmark}

\begin{abstract}
We present an overview of phase field modeling of active matter systems as a tool for capturing various aspects of complex and active interfaces. We first describe how interfaces between different phases are characterized in phase field models and provide simple fundamental governing equations that describe their evolution. For a simple model, we then show how physical properties of the interface, such as surface tension and interface thickness, can be recovered from these equations. We then explain how the phase field formulation can be coupled to various active matter realizations and discuss three particular examples of continuum biphasic active matter: active nematic-isotropic interfaces, active matter in viscoelastic environments, and active shells in fluid background.
Finally, we describe how multiple phase fields can be used to model active cellular monolayers and present a general framework that can be applied to the study of tissue behaviour and collective migration.\footnote{These are the lecture notes presented at the Initial Numerical Training on Active Matter in a training school for the MSCA-ITN ActiveMatter, (\url{http://active-matter.eu})}
\end{abstract}
\maketitle
\section{Introduction}
\begin{figure}[b]
\centering
  \includegraphics[width=1.0\linewidth]{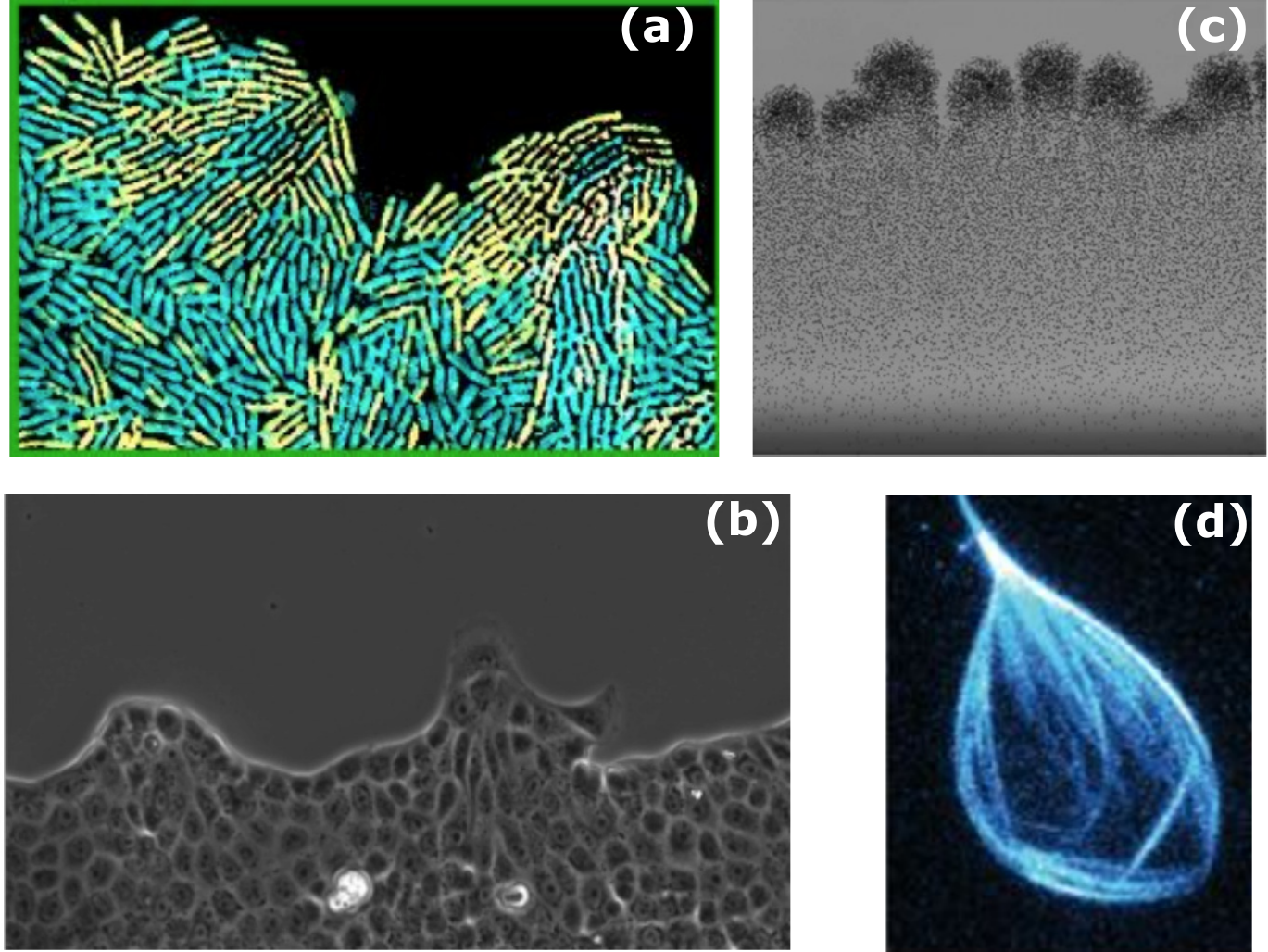}
  \caption{{\bf Examples of natural and synthetic active interfaces.} (a) Bacterial competition during {\it Pseudomonas ariginosa} biofilm invasion. Fast and slow moving bacterial strains are shown by yellow fluorescent protein (YFP) and cyan fluorescent protein (CFP), respectively. Figure adapted from~\cite{meacock2020bacteria}. (b) Epithelial cell progression during wound healing. Figure adapted from~\cite{doostmohammadi2015}. (c) Synthetic active matter composed of colloidal rollers forming fingering patterns at the progressive front. Figure adapted from~\cite{driscoll2017unstable}. (d) Self-deforming active droplet composed of microtubule-motor protein mixtures in a fluid background. Figure adapted from~\cite{keber2014topology}.}
  \label{fig:examp}
\end{figure}
While understanding pattern formation within active matter has taken important strides in the past decades~\cite{marchetti2013hydrodynamics,needleman2017active}, the vast majority of active systems are characterized by dynamic interfaces between an active phase and the surrounding medium, which can itself be active or passive with significantly different properties (Fig.~\ref{fig:examp}). Whether it is a bacterial biofilm invading its surrounding environment~\cite{burmolle2006enhanced,nadell2015extracellular,meacock2020bacteria}, a cellular tissue closing a wound~\cite{brugues2014forces}, or a cluster of human cancer cells migrating into extracellular matrices~\cite{friedl2009collective,friedl2012classifying}, the interaction between active matter and its environment plays a crucial role in determining the behavior of prominent physiological processes. 

In addition, even within the active matter itself, material properties are not always homogeneous and interfaces between active phases with different physical and chemical properties can arise. Striking examples include synthetic self-propelled particles that phase separate because of their different self-propulsion~\cite{zhang2020active}, competition between different phenotypes during bacterial collective invasion~\cite{meacock2020bacteria}, and cellular segregation in epithelial cells with distinct cell-cell interactions~\cite{balasubramaniam2020nature}. 

To correctly model such complex systems, it is then crucial to both capture the dynamics of each phase independently but also to accurately track the dynamic interface between them.
This can represent, however, a major modelling challenge.
For example, there are many relevant biophysical scenarios where active cellular materials interact with extracellular matrices with viscoelastic properties~\cite{nadell2015extracellular,chaudhuri2020effects}.
Phase field methods provide a generic and versatile framework that is adaptable to such complex situations and allow for the modelling of multiple active or passive phases with their own complex dynamics and interactions.

\section{Phase field description of diffuse interfaces\label{sec:phasefield}}

The phase field method, originally introduced to model solidification processes~\cite{RevModPhys.49.435,Fix1982PhaseFM,doi:10.1142/9789814415309_0005}, has since then seen a wide range of applications, such as the description of fingering and elastic surface instabilities, fluidization and crystallization in complex media, and the modeling of soft vesicles, see~\cite{doi:10.1080/00018730701822522} for a review.
The basic idea behind this method is to describe interfaces between two distinct phases implicitly using an auxiliary field $\phi$ whose value distinguishes the two phases.
For instance, $\phi=0$ can denote the first phase while $\phi=1$ the second, but this choice is arbitrary.
The phase field is then subject to a mixing free energy such that the balance between mixing and demixing effects results in two distinct stable phases with a diffusive interface of small, but finite, thickness.
Under complete demixing, a sharp interface separates these two phases, while the presence of a mixing free energy results in diffusion of the order parameter in the interfacial region.
This method is extremely versatile and can be used to describe interfaces between a variety of different phases, such as the solid and liquid phases of a melting material, the nematic and homogeneous phases of a biphasic liquid crystal, or simply the inside and outside of a vesicle or a cell.
In what follows, we present a simple phase field model with two phases and show how it can be coupled to flow and stresses.
 
\subsection{Mathematical basis}

\begin{figure}
 \centering
\includegraphics[scale=.8]{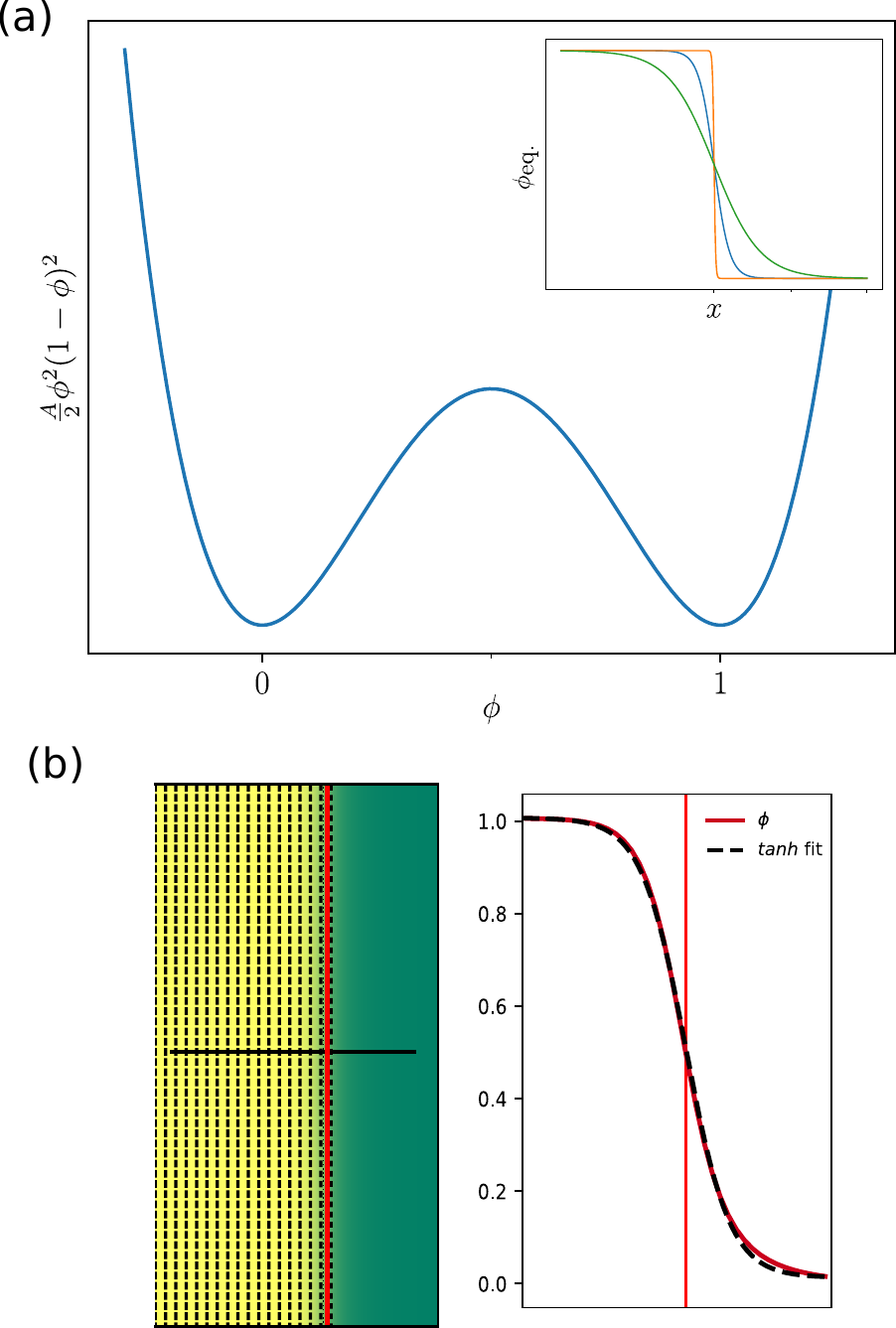}
\caption{{\bf Analysis of interfacial profile for $\phi$.} (a) The double-well potential corresponding to the demixing free energy in Eq.~\eqref{equation:CH free energy}. The inset shows the equilibrium phase field profile from Eq.~\eqref{eq:phieq} for different values of the interface width $\lambda$. The interface becomes sharp as $\lambda$ goes to $0$. (b) Simulation setup for measuring the interface between two phases with $\phi=1$ (yellow) and $\phi=0$ (green) in a complex fluid with nematic order parameter, see section~\ref{sec:actnem}. The phase field profile is measured along the black solid line and even in this it is very close to the theoretical $\tanh$ form of $\phi_{\text{eq.}}$.\label{fig:surface}}
\end{figure}

We present here the equations of a simple phase field model describing the demixing of two distinct phases and show analytically how the model parameters relate to the physical properties of the interface.
For simplicity, we assume that the phase field $\phi$ relaxes diffusively towards the minimum of the free-energy $\mathcal F = \mathcal F[\phi, \nabla \phi]$ according to the equation of motion
\begin{equation} \label{eq:phi}
    \partial_t \phi = -\frac{\delta {\mathcal{F}}}{\delta \phi} = - \frac{\partial {\mathcal{F}}}{\partial \phi} + \nabla \cdot \frac{\partial {\mathcal{F}}}{\partial \nabla \phi}.
\end{equation}
This equation of motion, which was first introduced in~\cite{ALLEN1972423} for the study of binary alloys, is arguably the simplest choice of evolution for the phase field $\phi$ but others are possible and can be found throughout the literature~\cite{doi:10.1080/00018730701822522}.
A simple free-energy exhibiting two distinct phases is the classical Cahn-Hiliard free energy first introduced in~\cite{doi:10.1063/1.1744102}, which can be written as
\begin{equation} \label{equation:CH free energy}
    \mathcal F_{\text{CH}} = \int \dd \vect x\, \left\{\frac{A}{2}\:\phi^2\left(1-\phi\right)^2 + \frac{\kappa}{2}(\nabla \phi)^2 \right\}.
\end{equation}
The first term is the demixing free energy and corresponds to a double well-potential with minima at $\phi=0$ and $\phi=1$ corresponding to the first and second phases, respectively, while the second term is the mixing part that penalises gradients $\nabla \phi$ (see Fig.~\ref{fig:surface}(a)).
Note the phase field $\phi$ is not conserved in equation~\eqref{eq:phi} but a conservation constraint can be easily introduced using Lagrange multipliers, see for example~\cite{jacqmin1999calculation,Yue2004}.

In order to get a better understanding of the parameters appearing in equation~\eqref{equation:CH free energy}, we search for steady state solutions of equation~\eqref{eq:phi} which leads us to solve
\[
    \frac{\delta \mathcal F_{\text{CH}}} {\delta \phi} = A \phi (1 - \phi) (1-2 \phi) - \kappa \Delta \phi = 0.
\]
It is easy to verify that
\begin{equation}
\phi_{\text{eq.}}(x) = \frac 1 2 \tanh \left( \frac{x_0 - x}{\sqrt{4 \kappa / A}} \right) + \frac 1 2,\label{eq:phieq}   
\end{equation}
is a solution of the above equation with boundary conditions $\phi_{\text{eq.}}(-\infty) = 1$ and $\phi_{\text{eq.}}(+\infty) = 0$.
This corresponds to a one-dimensional interface along the $x$-axis located at $x_0$ with interface width $\lambda = \sqrt{2\kappa / A}$, (see inset of Fig.~\ref{fig:surface}(a)).
Note that this solution is the absolute minimum of the free energy for these boundary conditions, but that infinitely many unstable minima exist where the phase field oscillates multiple times between $\phi=0$ and $\phi=1$~\cite{Mauri1996}.

Inserting $\phi_{\text{eq.}}$ into the free energy density, Eq.~\eqref{equation:CH free energy}, and integrating perpendicularly to the interface allows us to directly obtain the surface tension between the two phases $\phi=1$ and $0$ as
\begin{align*}
    \gamma &= \int_{-\infty}^{\infty} \dd x\, \left\{\frac{A}{2}\:\phi_{\text{eq.}}^2\left(1-\phi_{\text{eq.}}\right)^2 + \frac{\kappa}{2}(\nabla \phi_{\text{eq.}})^2 \right\} \\
    &= \sqrt{A \kappa} / 6.
\end{align*}

A similar analysis can be carried out in two dimension for the case of a radially symmetrical droplet even though no analytical solution for the profile exists in this case.
Considering a circular droplet whose radius $R$ is much larger that the interface width $\lambda$, it is easy to see that the Laplacian in radial coordinates can be approximated by $\Delta = \partial_r^2 + 1/r \partial_r \approx \partial_r^2 + \mathcal O(\lambda/R)$, in which regime the hyperbolic tangent profile from above is recovered.
When the radius of the droplet is somewhat similar to the interface width, this term can not be neglected and the interface profile will only be approximated by $\phi_{\text{eq.}}$.
Also note that in the non-stationary case $\partial_t \phi \neq 0$ the interface will also deviate from its equilibrium profile, but these deviations are found to be small in practice~\cite{Yue2004}.

Using a free energy-based description makes phase field models of interfaces easily adaptable for modeling different types of complex microstructured fluids, such as liquidc crystals or viscoelastic fluids, where the mixing free energy for the phase field is simply complemented by the free energy of the microstructured fluid~\cite{Yue2004}. Another advantage of the phase field models compared to the other approaches for modeling fluid interfaces is that it can be proven that the energy is always conserved even in the case of describing complex fluids such as liquid crystals~\cite{lin1995nonparabolic}.

\subsection{Coupling to hydrodynamics}

One attractive feature of phase models is that their governing equation for the binary order parameter $\phi$ can be easily coupled to hydrodynamic equations of fluid flow and other order parameters such as orientation field, polarity field, and even viscoelasticity. In this part we introduce a general framework through which such coupling is achieved, starting with a generic coupling to hydrodynamics and discussing specific examples from active nematics (where coupling to orientation field is introduced) and active matter in viscoelastic surrounding (where coupling to viscoelasticity is described). We further discuss how the generic framework for coupling to other order parameters can be adapted for more complex setups such as active shells within fluid backgrounds.

In order to couple phase field model to fluid flow, the governing equation for the binary order parameter $\phi$ is simply supplemented with an advective flux of $\phi$, $\vect{j}=\vect{u} \phi$, where $\vect{u}$ is the fluid velocity. This gives:
\begin{equation}
\label{eq:phiu}
\partial_t \phi + \vecnabla \cdot\vect{j} = -\frac{\delta {\mathcal{F}}}{\delta \phi}.
\end{equation}
As such, coupling to flow is easily achieved without introducing any additional parameters in the phase field equation. In its most general form, the fluid flow $\vec{u}$ is in turn determined from the Navier-Stokes equation:
\begin{equation}
\label{eq:u}
\rho\left(\partial_t \vec{u} + \vec{u}\cdot\vecnabla \vec{u}\right) = \vec{\nabla}\cdot\matr{\Pi},
\end{equation}
where $\rho$ is the fluid density and $\matr{\Pi}$ is the stress tensor. The stress tensor, in general, comprises fluid pressure $p$ and viscous stresses $\eta\matr{E}$, with $\eta$ the dynamic viscosity and $\matr{E}=(\vecnabla u)^{\text{Symm}}$ the strain rate tensor that characterises the symmetric part of the velocity gradient tensor. 

The presence of an interface between two phases gives rise to additional stresses in the flow field. To capture these additional stresses, the back-coupling from the binary order parameter $\phi$ to the fluid flow equation is introduced through capillary stresses~\cite{sulaiman2006lattice,blow2014biphasic,cates2018Theories,kempf2019active}:
\begin{equation}
\label{eq:capp}
\matr{\Pi}^{\text{capillary}} = \left(\mathcal{F}-\mu\phi\right)\id - \vecnabla \phi\frac{\partial\mathcal{F}}{\partial\vecnabla\phi},
\end{equation}
where $\mu=-\delta \mathcal{F}/\delta \phi$ is the chemical potential and $\id$ is the identity matrix. As such, the first term describes contributions to the fluid pressure (isotropic contributions to stress tensor) due to particle exchange between two phases through chemical potential. The second term on the right-hand-side of the Eq.~\eqref{eq:capp} characterizes anisotropic contributions to the fluid stress due to the presence of surface tension between the binary phases. Calculating the divergence of the capillary stress, it can be shown that the corresponding force field simplifies to~\cite{sulaiman2006lattice}
\begin{equation}
\label{eq:Fcapp}
\vect{F}^{\text{capillary}}=\vecnabla\cdot\matr{\Pi}^{\text{capillary}}=\phi\vecnabla\mu.
\end{equation}
Therefore, from Eqs.~\eqref{eq:phi}-\eqref{eq:Fcapp}, the coupled set of equations for a phase field model in the presence of hydrodynamic fluid flow can be written as:
\begin{eqnarray}
\label{eq:phi-u}
\partial_t \phi + \vecnabla\cdot\left(\vect{u}\phi\right) &=& \mu, \nonumber\\
\rho\left(\partial_t \vect{u} + \vect{u}\cdot\vecnabla \vect{u}\right) &=& \vecnabla \cdot\matr{\Pi}+\phi\vecnabla \mu,
\end{eqnarray}
where, as before, the stress tensor $\matr{\Pi}$ includes pressure and viscous contributions. 

Eq.~\eqref{eq:phi-u} can be solved numerically using hybrid schemes combining a finite-difference integration method~\cite{leveque2007finite} for the scalar binary order parameter with the desired fluid solver, e.g. lattice Boltzmann method~\cite{Marenduzzo2007,desplat2001ludwig} (see~\cite{carenza2019lattice} for a recent review of lattice Boltzmann method for simulating active fluids). 

The framework discussed above represents a general formulation for the coupling between the binary order parameter and hydrodynamic fluid flow and can be easily adapted to problems where the fluid flow is generated by active particles. For active fluids, additional active stresses can easily be added to the stress tensor~\cite{cates2018Theories,julicher2018hydrodynamic,doostmohammadi2018active} and their exact form depends on the type of the active system under consideration. The important point, however, is that the effect on the evolution of the binary order parameter can be described by only considering the active flux ($\vect{j}=\vect{u}\phi$) that is solely determined from the resulting velocity field. In what follows, we discuss specific examples from active nematic models to make this more clear.

\section{Continuum biphasic active systems}
\begin{figure}[h]
\centering
  \includegraphics[width=1.0\linewidth]{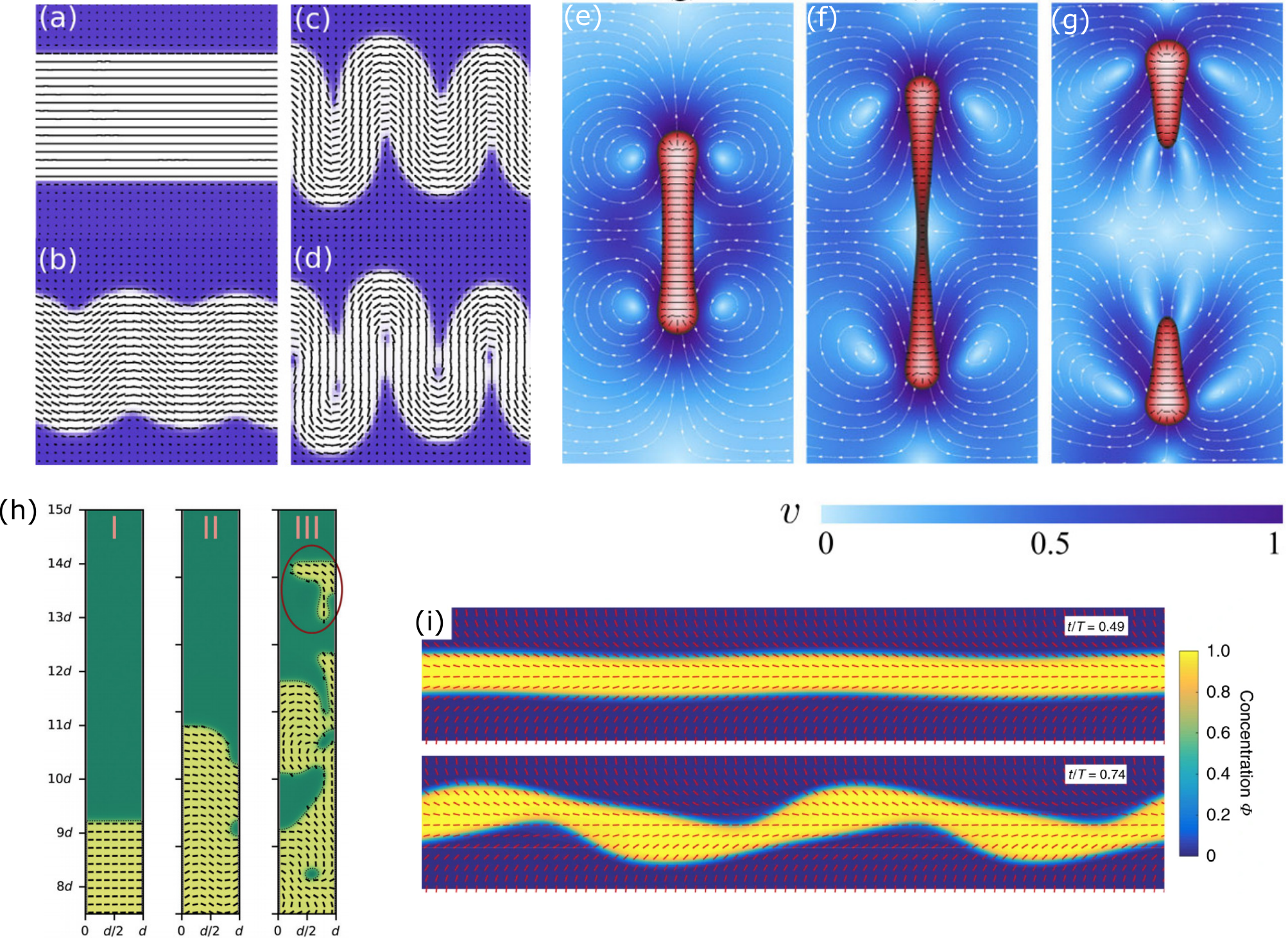}
  \caption{{\bf Modeling active biphasic systems composed of active nematics and isotropic phases.} (a-d) Temporal evolution of the instability of an active nematic band (white region) within an isotropic fluid background (blue region). The black solid lines represent the director field of the active nematic. An initially flat interface (a) is unstable to bend deformation of the active nematics (b), which grow in amplitude (c) and generate a highly deformed interface (d). Figure adapted form~\cite{blow2014biphasic}. (e-f) A drop of active nematic elongates (e), forms a narrow neck (f), and eventually divides (g). The black solid lines represent the director field of the active nematic and white vectors indicate the flow streamlines. The active nematic region is marked as red, while the isotropic fluid region is shown in blue. Figure adapted form~\cite{giomi2014spontaneous}. (h) Growing active nematic (yellow region) invading a surrounding isotropic fluid (green region). I-III show different patterns of invasion for increasing activities. Figure adapted form~\cite{kempf2019active}. (i) Active nematic fluid (yellow region) in a background passive liquid crystal develops instabilities that do not grow beyond a certain length set by the properties of the background liquid crystal. Red solid lines represent the director field of the background liquid crystal that are initially patterned to form a splay region in the middle. Figure adapted form~\cite{turiv2020polar}.}
  \label{fig:actnem}
\end{figure}

\subsection{Active nematic interfaces\label{sec:actnem}}

Active nematics represent a class of active materials that are characterized by having a local orientational order parameter~\cite{cates2018Theories,julicher2018hydrodynamic,doostmohammadi2018active}. This comprises solutions of elongated particles such as bacteria~\cite{volfson2008biomechanical,meacock2020bacteria}, microtubule filaments put in motion by kinesin motor proteins~\cite{Sanchez2012,guillamat2016control} or actin filaments and myosin mixtures~\cite{kumar2018tunable}, but also spindle shaped cells such as mouse embryonic fibroblasts~\cite{duclos2018spontaneous} and neural progenitor cells~\cite{kawaguchi2017topological}.
More surprisingly, deformable cells such as Madine-Derby Canine Kidney cells (MDCK)~\cite{saw2017topological}, human bronchial cells (HBC)~\cite{blanch2018turbulent}, and human fibrosarcoma~\cite{yashunsky2020chiral} can also be described by active nematic theories, where the orientational order is defined as the direction of cell elongation.
In nematic systems, the orientational order is characterized by a second-rank nematic tensor $\matr{Q}$.
The field associated with the orientational order of elongated particles must be represented by a tensor since it must show head-tail symmetry, i.e. be apolar. For this reason, it can not be defined using a directional unit vector $\uvect{n}$ only and instead higher moments must be used to define the nematic tensor ${\matr Q} = q\frac{d}{d-1}(\hat{n}^\transp\hat{n}-\id/d)$, where $q$ is the magnitude of the nematic order that corresponds to the largest eigenvalue of $\matr{Q}$ (see~\cite{doostmohammadi2018active} for a recent review of active nematics). 

It is well-established that for an active nematic system, the prominent contribution to active stresses is proportional to the nematic tensor ${\matr Q}$ such that the active stress takes the form $\zeta{\matr Q}$~\cite{voituriez2005spontaneous,ramaswamy2010mechanics,Marchetti13,doostmohammadi2018active}, where $\zeta$ determines the strength of activity. Therefore, in order to account for the effect of activity, an active stress proportional to the nematic tensor ${\matr Q}$ is added to the stress tensor is Eq.~\eqref{eq:phi-u}: $\matr{\Pi}^{\text{active}}=\phi\zeta{\matr Q}$. Since the divergence of the stress determines the force acting on the fluid, this means that any gradients in the nematic tensor ${\matr Q}$ result in additional forces and actively generate fluid flow~\cite{ramaswamy2010mechanics}. Moreover, the proportionality of the stress to the binary order parameter $\phi$ ensures that the active stress is only applied in the phase with $\phi=1$ and is zero in the other phase where $\phi=0$. Such a form for the active stress can be used to model systems with a binary mixture of active and passive nematics (where both phases have orientational order, but only one of them is active), such as mixtures of live and dead bacteria~\cite{patteson2018propagation} or a mixture of motor proteins-filaments in which only a part of the mixture is activated through light-activated motor proteins~\cite{zhang2019structuring}. 

In several experimental conditions, however, an active phase of particles with orientational order, forms an interface with an otherwise isotropic fluid. This can be easily accommodated within the framework described here by ensuring that the passive phase $\phi=0$ has no orientational order, i.e., ${\matr Q}\rightarrow{\matr 0}$ when $\phi\rightarrow 0$. To this end, the Landau-De Gennes free energy for the nematic $\mathcal{F}_{\text{L-G}}=\frac{1}{2}{\matr Q}^2+\frac{1}{4}{\matr Q}^4$~\cite{DeGennes,BerisBook}, is coupled to the binary order parameter $\phi$ as follows:
\begin{equation}
\label{eq:phi-Q}
\mathcal{F}_{\text{L-G}}=\frac{1}{2}\phi{\matr Q}^2+\frac{1}{4}{\matr Q}^4.
\end{equation}
This free energy ensures that in the active nematic phase $\phi=1$, the orientational order is retained, while in the passive phase $\phi=0$, the magnitude of orientational order that minimizes the free energy --- i.e. ${\matr Q}$ for which $\delta\mathcal{F}/\delta{\matr Q}=0$ --- diminishes to zero, resulting in an isotropic fluid~\cite{blow2014biphasic,kempf2019active}. In Figure~\ref{fig:surface}b, we compare analytical results for the one-dimensional $\phi$ profile to the results from numerical integration of Eq.~\eqref{eq:phi}.

An interesting feature of active nematic interfaces compared to their passive counterpart, is the emergence of the phenomenon of {\it active anchoring}~\cite{blow2014biphasic}: in the absence of any anchoring energies, the presence of activity alone results in a certain alignment of active particles at the interface. This effect has been shown for dividing bacteria, where cell division provides a source of active force generation, that align parallel to the interface of their colony~\cite{doostmohammadi2016defect} and can also be seen in Fig.~\ref{fig:examp}(a). Considering the active force
\begin{align}\label{equation:activeforce}
    F_\alpha^{\text{active}}=&-\zeta\partial_\beta\left(\phi Q_{\alpha\beta}\right) \nonumber \\
    =&-\zeta\left(\partial_\beta\phi S\right)\left(2n_\alpha n_\beta-\delta_{\alpha\beta}\right) \nonumber \\
    &-2\;\zeta\phi S\left(n_\alpha\left(\partial_\beta n_\beta\right)+\left(\partial_\beta n_\alpha\right)n_\beta\right),
\end{align}
the active anchoring can be understood by decomposing the active force parallel and perpendicular to the interface between active and passive phases~\cite{blow2014biphasic}:
\begin{align}
    f_\perp&=m_\alpha f_\alpha=\zeta\left\lvert\nabla\left(\phi S\right)\right\rvert\left(2\left(m_\alpha n_\alpha\right)^2-1\right),\\
    f_{\parallel}&=l_\alpha f_\alpha=2\;\zeta\left\lvert\nabla\left(\phi S\right)\right\rvert\left(l_\alpha n_\alpha\right)\left(m_\beta n_\beta\right),
\end{align}
where $\vec{m}\equiv-\vecnabla S/\left\lvert\vecnabla S\right\rvert\equiv-\vecnabla\left(\phi S\right)/\left\lvert\vecnabla\left(\phi S\right)\right\rvert$ and $\vec{l}$ define the outward surface normal and the surface tangent directions, respectively.

Depending on the sign of the activity parameter $\zeta$, the normal force on the interface $f_\perp$, points outwards ($\zeta<0$) or inwards ($\zeta>0$) for $\vec{n}\perp\vec{m}$ and hence induces active forces pushing the interface outwards or inwards, respectively. Therefore, for $\zeta>0$ activity favors alignment of active particles parallel to the interface, while $\zeta<0$ favors perpendicular alignment. The case of $\zeta > 0$ describes extensile active systems such as bacteria, where particles pull the surrounding fluid from their side and push it along their tail and head. The $\zeta < 0$ case, on the other hand, corresponds to contractile active systems such as actomyosin networks, where active constituents contract along their direction of elongation.

The method presented here has been extensively used to model fundamental aspects of active interfaces, including the study of the instabilities of active nematic interfaces~\cite{blow2014biphasic,blow2017motility} (Fig.\ref{fig:actnem}(a-d)), describing dynamics of elongation, motility, and division of active nematic droplets as a model of eukaryotic cells~\cite{giomi2014spontaneous,carenza2019chaotic} (Fig.\ref{fig:actnem}(e-g)), and wetting dynamics of active nematics on solid surfaces~\cite{coelho2020propagation,neta2020wetting}. Furthermore, the biphasic modeling of active nematic interaction with isotropic fluid has been used to describe morphologies of growing bacterial colonies~\cite{doostmohammadi2016defect,dell2018growing}, active matter invasion into new territories~\cite{kempf2019active} (Fig.\ref{fig:actnem}(h)), and bacterial streams within pre-patterned medium in living liquid crystals~\cite{turiv2020polar} (Fig.~\ref{fig:actnem}(i)). 

\subsection{Active matter in a viscoelastic medium}
\begin{figure}[h]
\centering
  \includegraphics[width=0.8\linewidth]{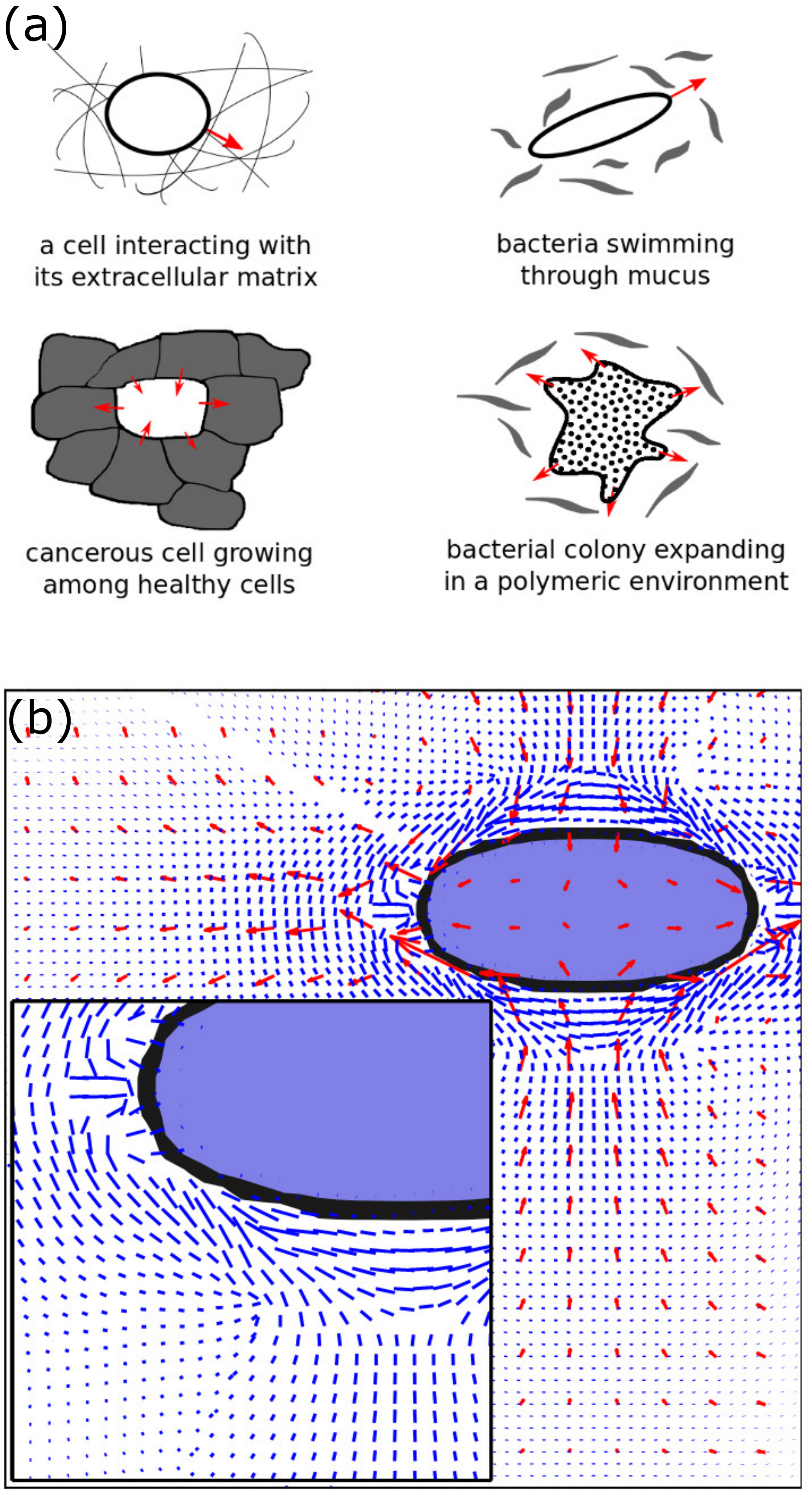}
  \caption{{\bf Modeling active biphasic system composed of active nematics and passive polymeric phases.} (a) A schematic of various interpretations of the two-phase active-viscoelastic model. (b) Typical flow field of a drop of active nematics elongating during the initial steps of cell division. Velocity vectors are
shown as red arrows, and polymer deformations are shown as blue line segments. Inset: Zoom of the lower left part of the cell, showing polymers aligning with the stretching direction. Figure adapted from~\cite{plan20}.}
  \label{fig:visco}
\end{figure}
The exact same procedure used in the previous section can be followed to introduce couplings to other order parameters. One important example is active matter in a viscoelastic environment which has applications to the modelling of several biophysical systems of interest.
For example, cells in extracellular matrices are characterized by the cross talk between an active phase and a medium which has viscoelastic properties~\cite{chaudhuri2020effects} (Fig.\ref{fig:visco}(a)). Following the framework introduced in the previous section, we extend our formalism to the case of an active nematic in contact with a fluid phase that, rather than being isotropic, has now viscoelastic properties. 

To characterize a viscoelastic fluid, it is common to introduce a conformation tensor ${\matr C}$ as the corresponding order parameter, which describes the orientation and elongation of polymers that are present in the fluid~\cite{hemingway2015active,plan20}. The conformation tensor $\matr C$ is defined such that: (i) it satisfies ${\matr C}=\id$ at equilibrium, (ii) its trace $\tr[{\matr C}]$ characterizes the square of the polymer elongation, and (iii) the eigenvector corresponding to its largest eigenvalue represents the local polymer orientation. Therefore, as before, a coupling between the binary order parameter and the conformation tensor can be introduced such that within the active phase ($\phi=1$) there is an orientational order characterized by the nematic tensor ${\matr Q}$ but no viscoelasticity ${\matr C}={\matr 0}$, while in the passive phase ($\phi=0$) there are no active nematic particles ${\matr Q}={\matr 0}$ but polymers are present ${\matr C}\neq {\matr 0}$.
To ensure this, the free energy of polymers is coupled to the binary order parameter $\phi$ as follows:
\begin{equation}
\label{eq:phi-Q}
\mathcal{F}_{\text{polymer}}=\frac{\nu}{\tau}\frac{1}{2}\left(1-\phi\right)\left(\tr[{\matr C}-\id]-\log\left(\det[{\matr C}]\right)\right),
\end{equation}
where $\nu$ characterizes the polymer viscosity and $\tau$ is the polymer relaxation time. Note that the factor $1-\phi$ in front of the polymer free energy ensures that the order parameter is retained within the passive phase $\phi=0$, while it diminishes otherwise when $\phi=1$.
Finally, since polymers are within the passive phase, no extra addition to active stresses is required. However, passive stresses need to be accounted for in Eq.~\eqref{eq:phi-u} as follows:
\begin{equation}
\label{eq:C}
{\matr \Pi}^{\text{polymer}}=\frac{\nu}{\tau}(1-\phi)({\matr C}-\id).
\end{equation}

It is noteworthy that the formulation described above uses an Oldroyd-B model of polymer relaxation since it is one the simplest and most widely-used polymer models~\cite{plan20}. The framework can be trivially extended to other constitutive models of viscoelastic fluids in two and three dimensions. 

The important feature of the present framework that couples viscoelasticity, fluid flow, and activity through the phase field formulation is its versatility, which means that it can be applied to various biophysical examples. For instance, similar formulation has been used to study a model of cell division and motility within polymer gels with variable viscoelastic properties (Fig.\ref{fig:visco}(b)), and to study the hampering effect of polymers on generic instabilities of active matter in a viscoelastic environment~\cite{plan20} (Fig.~\ref{fig:visco}). 

\subsection{Self-deforming active surfaces in a fluid background}

\begin{figure}[h]
\centering
  \includegraphics[width=1.0\linewidth]{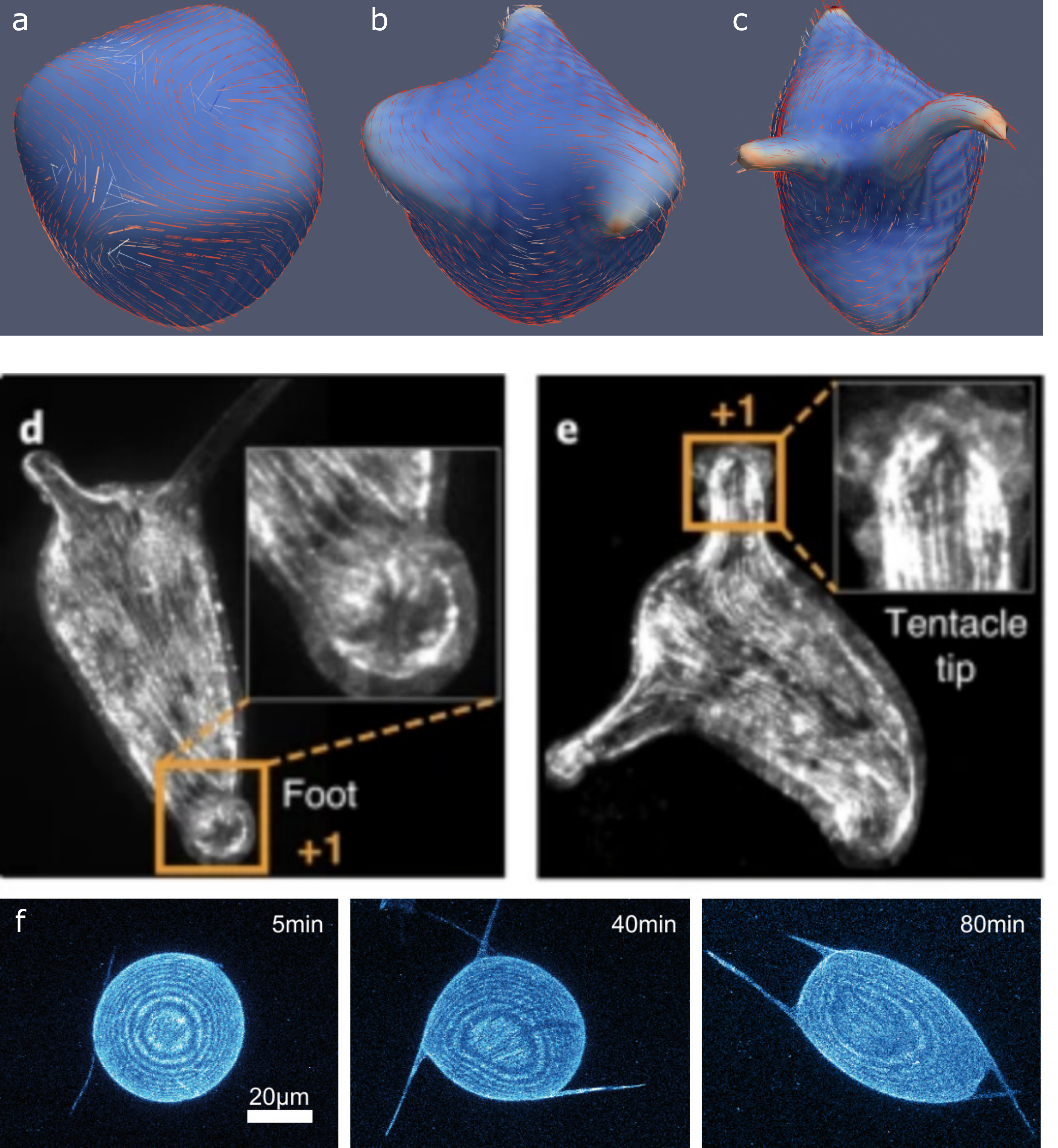}
  \caption{{\bf Modeling three-dimensional active self-deforming surfaces.} (a-c) Temporal evolution of an active shell made of active nematics in an isotropic fluid background. Active shell develops protrusions at the position of $+1/2$ defects. The protrusions eventuality develop $+1$ defects at their tip leaving behind $-1/2$ defects. The solid lines illustrate the nematic director and are colored by the magnitude of the nematic order. The shell is colored by the curvature of the surface. Figure adapted from~\cite{metselaar2019topology}. (d-e) Developments of protrusions and topological defects during the morphogenesis of regenerating animal {\it hydra}. Figure adapted from~\cite{maroudas2020topological}. (f) Temporal evolution of protrusions in a shell of active nematics build from microtubule-motor protein mixtures. Figure adapted from~\cite{keber2014topology}.}
  \label{fig:shell}
\end{figure}

So far, we have described applications of the phase field method to binary mixtures, where $\phi=0$ is used to characterize one phase and $\phi=1$ to describe the other, with $\phi=0.5$ marking the interface between the two phases. It is, however, possible to extend this framework to model scenarios in which thin shells of one phase are embedded within another phase. An important example is the actomyosin cortex of eukaryotic cells that plays a vital role in shaping cell deformation and various cell functions~\cite{naganathan2014active}. Another interesting example is the active deformable shell formed by stabilizing microtubule-motor protein mixtures at the oil-water interface of a droplet~\cite{keber2014topology,zhang2016dynamic,guillamat2018active}. The resulting active shell continuously exerts active stresses on the surrounding medium, generating dynamic patterns of motion and exotic morphologies that can be considered as examples of self-deforming and self-shaping materials (see Fig.~\ref{fig:shell}(a)). 

The distinguishing feature of these systems is that the shell itself is active and as such activity is generated at a thin interface between the inside and the outside of the shell. The phase field model of active interfaces described in this section can be easily adapted to model such as active shells by describing the coupling between the binary order parameter $\phi$ and the desired order parameter (in the case of active nematic this is the nematic tensor ${\matr Q}$ which is the source of active stresses) such that the active phase is only stabilized at the interface, i.e., where  $\phi \approx 0.5$. For instance, starting from the active nematic system described in Sec.~\ref{sec:actnem}, it is only required to modify the free energy as follows:
\begin{equation}
\label{eq:phi-Qn}
\mathcal{F}_{\text{L-G}}=\frac{1}{2}\left(1-2\abs{\phi-\frac{1}{2}}\right){\matr Q}^2+\frac{1}{4}{\matr Q}^4,
\end{equation}
where the term $\left(1-2\abs{\phi-\frac{1}{2}}\right)$ ensures that ${\matr Q}\to {\matr 0}$ as both $\phi \to 0$ and $\phi \to 1$, while the orientational order is retained at the interface ($\phi \approx 0.5$). Again, this provides only a simple example of how such coupling can be achieved. The framework can be trivially extended to consider further complexities of the interface, for example, by including Helfrich type free energies that describe bending rigidity of the shell~\cite{metselaar2019topology}.

This method has been recently applied to model self-deforming active nematic shells in fluid backgrounds, reproducing topological defect dynamics on the shell surface observed in the experiments on microtubule-kinesin motor mixtures~\cite{keber2014topology,zhang2016dynamic,metselaar2019topology}, and predicting exotic morphologies that can be obtained by tuning the activity of the shell. More importantly, such modeling provides a tool to probe the nature of cross talk between activity and self-induced surface curvature. For example, it is predicted that protrusions are initiated at the locations of $+1/2$ defects in the orientation field, creating finger-like structures with a $+1$ topological defect at their tip, leaving behind $-1/2$ topological defects in regions with negative surface curvature (Fig.~\ref{fig:shell}(b)). Similar dynamics and cross talk between activity and morphology has been recently reported in the development of regenerating animal {\it hydra} where actin filaments form an active shell that is capable of self-deformation~\cite{maroudas2020topological} (Fig.~\ref{fig:shell}(c)).

The examples described above provide an introduction to the applications of phase field model for modeling active systems.
We focused particularly on active nematic interfaces, where dynamics of phase field is coupled to a tensor order parameter and which describes orientational order of active particles. Importantly, phase field models have been applied to various realizations of active systems including scalar phase field models describing the phenomenon of motility-induced phase separation~\cite{wittkowski2014scalar} (see~\cite{cates2018Theories} for a recent review), as well as polar active matter, where phase field dynamics is coupled to a vector order parameter describing the polarity of active particles~\cite{negro2019rheology,carenza2019lattice}. The latter has been particularly successful in modeling cellular motility, where polar active matter is used to describes the dynamics of actomyosin networks inside the cell, governing cell deformation and its morphology~\cite{ziebert2011model,tjhung2012spontaneous,marth2016collective,carenza2019chaotic}

\section{Phase field modeling of cell monolayers}

In the previous sections we discussed applications of phase field models to continuum representations of active interfaces. Interestingly, the same framework can be adapted to agent-based modeling of active systems where each agent is represented by an individual phase. An important example is the phase field modeling of cell monolayers which allows to model the deformations of individual cells and describe their physical properties, as well as introducing cell-cell interactions easily through the interfaces of the individual phase fields. In this section, we discuss how such models are constructed and applied to the understanding of the dynamics of cellular layers.

\subsection{Cells as active deformable droplets}
\begin{figure}[b]
\centering
  \includegraphics[width=1.0\linewidth]{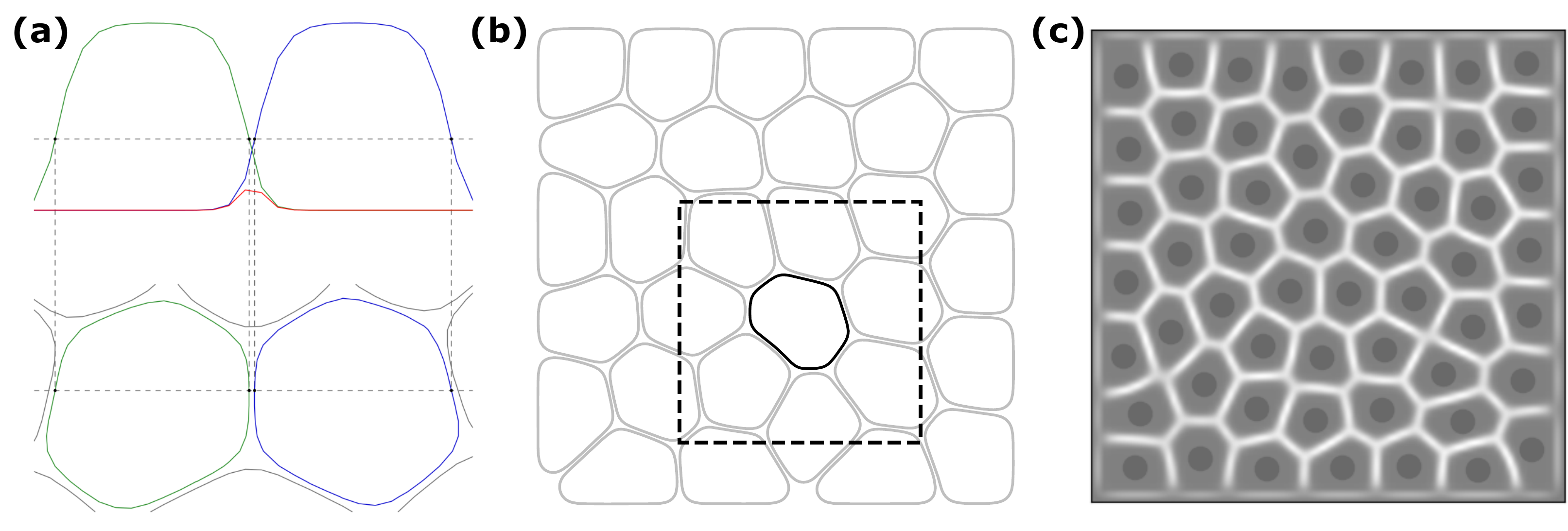}
  \caption{{\bf Modeling cell monolayer composed of multiple active phase fields representing each cell.} (a) Schematic representation of the phase fields of two cells (green and blue) as well as their overlap (red) seen from the side. (b) A phase field representation of a cell monolayer as a collection of active deformable drops seen from the top. The square around the cell represent the sub-domain on which the equations of motion are solved, see section~\ref{sec:cells:simulations}. (c) A confluent monolayer in confinement where the overlap between cells and their centres of mass are shown instead of the cell contours.}
  \label{fig:cells}
\end{figure}
{\it In silico} models of cell motility have an important role to play in unravelling the interplay between single cell properties and their collective dynamics~\cite{bi2015density,alert2020physical}.
There exist a breadth of numerical approaches to tackle this problem, such as cellular Potts models~\cite{graner1992simulation}, vertex models~\cite{farhadifar2007influence,bi2016motility,sussman2018anomalous}, continuum models~\cite{prost2015active,julicher2018hydrodynamic}, and phase field models~\cite{lober2015collisions}, see the extensive reviews~ \cite{alert2020physical,camley2017physical,shaebani2020computational}.
The phase field approach is particularly interesting as it allows to describe the collective behaviour of cells while modelling their physical properties at the individual level as well as their interactions explicitly.
It has been widely applied to problems involving single cells~\cite{lober2014modeling,ziebert2011model,ziebert2013effects,tjhung2012spontaneous}, few migrating cells \cite{PhysRevE.93.052405}, as well as systems of colliding binary cells~\cite{camley2014polarity,lober2015collisions}.
More recent works~\cite{marth2016collective,palmieri2015multiple,peyret2019sustained,wenzel2019topological,mueller2019emergence,loewe2019solid} have concentrated on the collective dynamics of a large number of cells and have uncovered interesting phenomena such as oscillatory patterns of cells under confinement~\cite{peyret2019sustained}, as well as solid-liquid and flocking transitions~\cite{giavazzi2018flocking}.
However, since this approach is computationally quite intensive, it is necessary to reduce the description of cells to their bare minimum by abstracting away their internal machinery while retaining their relevant physical properties.
Here we concentrate on a simple model that represents cells on a substrate in two-dimensions and assumes simple physical properties for the cells.

A fruitful approach first pioneered by \cite{palmieri2015multiple}, is to separate the description of the physical properties of the cells from their dynamics.
In this model, each cell is described as a deformable active particle whose shape is defined using an individual phase field (Fig.~\ref{fig:cells}(a-b)), while its velocity is described separately and is given by a force-balance equation.
A cellular monolayer consisting of $N$ cells can then be described by the equations of motion
\begin{equation}
\label{eq:cells:dynamics}
\partial_t \phi_i + \vect{v}_i \cdot \vecnabla \phi_i = - \frac{\delta {\mathcal{F}}}{\delta \phi_i}, \qquad i = 1, \ldots, N,
\end{equation}
where ${\mathcal F}$ is a free energy describing the physical properties of the cells, and $\phi_i$ and $\vect v_i$ are the phase field and the center-of-mass velocity of cell $i$, respectively.
The phase fields are defined such that $\phi = 0$ denotes the exterior of the cell $i$ while $\phi = 1$ denotes its interior.
This mirrors Eq.~\eqref{eq:phi} with the difference that there is a different phase field for each cell and that each individual cell and is driven by a simple advection term.
The velocity $\vect v_i$ for each cell can be obtained from a force balance equation: since Reynolds numbers are typically of the order of $\sim 10^{-4}$ for cell monolayers~\cite{kim2013propulsion}, we can safely assume overdamped dynamics and write
\begin{equation}
\label{eq:cells:force_balance}
\xi \vect{v}_i =\vect{F}^{\text{tot.}}_i = \vect{F}^{\text{pol.}}_i + \vect{F}^{\text{inter.}}_i,
\end{equation}
where $\xi$ is a friction coefficient and $\vect{F}^{\text{tot.}}_i$ is the total force exerted on cell $i$, which will be defined below.
Note that $\vect v_i$ is a single vector for each cell and does not depend on $\vect x$.
The definition of the total force $\vect{F}^{\text{tot.}}_i$ depends on the model and can include contributions from different sources such as passive interface forces, interactions with other cells or walls, or active forces such as polar or nematic driving.
Note that we have assumed here for simplicity that the free energy is the same for all cells but this does not need to be the case in general and it is straightforward to extend the above equation to cases where different cells have  different physical properties, see for example~\cite{balasubramaniam2020nature}.
Equations~\eqref{eq:cells:dynamics} and~\eqref{eq:cells:force_balance} are our master equations and define a general framework where only the free-energy $\mathcal F$ and the total force $\vect{F}^{\text{tot.}}_i$ on each cell need to be specified.

\subsection{Minimal model of cell monolayers\label{sec:cells FE}}

Let us now define a minimal model that allows the successful description of cell monolayers following \cite{palmieri2015multiple,mueller2019emergence} and write the free energy as a sum of three distinct contributions as $\mathcal{F} = \mathcal{F}_{\text{CH}} + \mathcal{F}_{\text{area}} + \mathcal{F}_{\text{rep.}}$.
The first term is the Cahn-Hilliard free energy~\eqref{equation:CH free energy}, which we rewrite as
\begin{equation}
\mathcal{F}_{\text{CH}} = \sum_i \frac \gamma \lambda \int \dd\vect x \left\{ 4 \phi_i^2 (1-\phi_i)^2 + \lambda^2 (\nabla \phi_i)^2 \right\},
\end{equation}
using the definitions of the interface width $\lambda = 2 \sqrt{\kappa/A}$ and surface tension $\gamma = \sqrt{A \kappa}/6$ from sec.~\ref{sec:phasefield} (up to an overall factor of $3$).
As shown in sec.~\ref{sec:phasefield}, this term creates and stabilises an interface between the interior ($\phi_i = 1$) and the exterior ($\phi_i = 0$) of each cell and results in a equilibrium shape for the interface that follows an approximate hyperbolic tangent profile of size $\lambda$.
The second term is a soft area constraint defined by
\begin{equation} \label{eq:Farea}
\mathcal{F}_{\text{area}} = \sum_i \mu \Big( 1 - \frac{1}{\pi R^2}\int \dd\vect x\, \phi_i^2 \Big)^2,
\end{equation}
which is a square potential of strength $\mu$ ensuring the cells areas $A_i = \int \dd x\, \phi_i^2$ are close to $\pi R^2$~\footnote{The area is proportional to the \emph{square} in order to ensure that it is always positive even when the phase field is slightly negative which can happen during simulations.}.
Note that the phase field $\phi_i$ is not conserved in Eq.~\eqref{eq:cells:dynamics} and that even though cells are mostly incompressible in three dimensions, the area that individual cells occupy within a monolayer can dramatically change as they are squeezed by their neighbours and expand in the direction perpendicular to the substrate~\cite{deforet2014emergence}.
The final term discourages overlap between cells and is simply given by
\begin{equation}
\mathcal{F}_{\text{rep.}} = \sum_i \sum_{j \neq i} \frac{\kappa}{\lambda} \int \dd \vect x \, \phi_i^2 \phi_j^2\, ,
\end{equation}
where $\kappa$ is the strength of the repulsion.
Note that this term implements a coupling between the cells and is proportional to the cell overlaps, see an illustration on Fig.~\ref{fig:cells}(c).
The physical properties of the cells are parameterized by $\lambda$, $\gamma$, $\mu$, and $\kappa$ which set the interface width, the surface tension, the strength of the elastic restoring force for the area, and the repulsive force between cells, respectively.
Normalisation is chosen such that properties of the cells are roughly preserved when the interface width $\lambda$ is rescaled, see~\cite{mueller2019emergence}.

Cellular deformations such as changes of shape or area should lead to forces at the cell boundaries and contribute to the overall force balance.
Such interface forces can be constructed in a thermodynamically consistent way using the free energy as
\begin{equation} \label{eq:cells:passive}
    \vect F_i^{\text{inter.}} = \int \dd \vect x \sum_j \left( \frac{\delta \mathcal F_{\text{rep.}}}{\delta \phi_j} - \frac{\delta \mathcal F_{\text{CH}}}{\delta \phi_j} - \frac{\delta \mathcal F_{\text{area}}}{\delta \phi_j} \right)\vecnabla \phi_i,
\end{equation}
see~\cite{palmieri2015multiple,mueller2019emergence} for more details.
Since $\vecnabla \phi_i$ is only non-zero at the cell boundary and pointing towards the cells center, this expression can be interpreted as the integral over the interface of the cell $i$ of the total force density generated by changes of the free-energy.
Contributions from interactions with other cells come with a positive sign (repulsion) while self contributions come with a negative sign (restoring forces).
With this and in the absence of any active contribution, a confluent cell monolayer will relax to a minimum of the total free energy and favour hexagonal lattice arrangement~(Fig.~\ref{fig:cells}(c)).

The great strength of this formulation is its versatility: it can be easily extended to more complex models that can include forces such as adhesion between cells~\cite{Zhang2020} or viscous friction~\cite{peyret2019sustained}.
Moreover, as shown in~\cite{peyret2019sustained}, it is also easy to model non-trivial boundary conditions (such as walls) by introducing a static phase field $\phi_{\text{walls}}$ that interacts with the cells with a simple repulsion term.

\subsection{Efficient simulation algorithm\label{sec:cells:simulations}}

The coupled system of differential equations defined by~\eqref{eq:cells:dynamics} and~\eqref{eq:cells:force_balance} can be solved using standard finite difference schemes~\cite{leveque2007finite}.
The main difficulty is the large number of phase fields $\phi_i$ that are required when simulating a monolayer with many cells.
In fact, it is easy to see that a naive simulation will have at least quadratic computational and memory requirements with respect to the number of cells due to each phase field extending over the whole domain.
This can be mitigated by realising that the individual phase fields are non-zero only in a well-defined region around the center of the corresponding cell, which allows to simulate each phase field only in a sub-domain centered around its center-of-mass~\cite{10.1371/journal.pone.0033501,mueller2019emergence}, see Fig.~\ref{fig:cells}(b).

A common choice is to select a square of lattice points that tracks the center of mass of each cell and use Dirichlet boundary conditions at the boundary of the sub-domain~\cite{loewe2019solid}.
It is important to note, however, that because the centre of mass of the cell can lie between lattice points, the location of the sub-domain changes abruptly every time the cell center moves by a lattice length along any axis.
This can lead to numerical artifacts when the size of sub-domain is not large enough compared to the radius of the cell.
A more refined approach is to use periodic boundary conditions instead~\cite{mueller2019emergence}, which has the advantage of being completely insensitive to the discrete hopping of the sub-domain and allows smaller sub-domain sizes compared to Dirichlet boundary conditions.

With such techniques, both the computational and memory costs of the algorithm scale linearly with respect to the number of cells when simulating a multicellular monolayer.
Note that even though cells are only simulated around their center of mass, the computation of the total force acting on a cell interface will still require the computation of some quantities over the whole domain, such as the sum of all phase fields $\sum_i \phi_i$ and their squares $\sum_i \phi_i^2$, but the number of such quantities is fixed and does not scale with the number of cells.
An implementation by the authors of the model presented in section~\ref{sec:cells FE} using these techniques can be found online at \url{https://github.com/rhomu/celadro}.

\subsection{Active polar driving}
\begin{figure}[h]
\centering
  \includegraphics[width=1.0\linewidth]{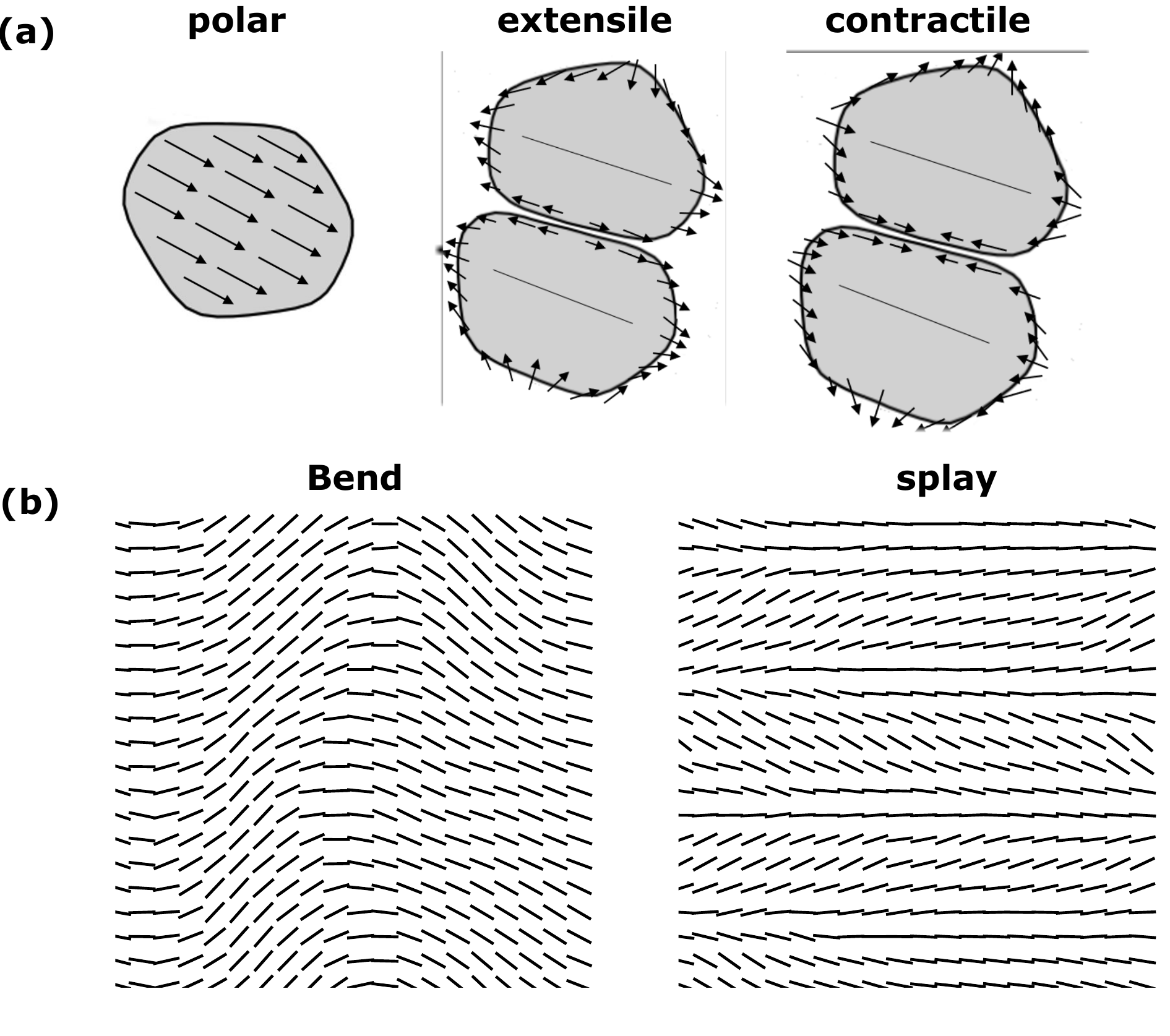}
  \caption{{\bf Active forces and instabilities in cell monolayer.} (a) Schematic representation of different sources of active force. Figure adapted from~\cite{Zhang2020}. (b) Coarse-grained nematic field of the cells showing the emergence of multicellular scale bend instability for extensile active forces and splay instability for contractile active forces acting at the individual cell level. The cells are driven out of equilibrium by an internal nematic degree of freedom.}
  \label{fig:tissue}
\end{figure}
The driving forces behind the movements of single cells on a flat substrate are well understood.
Directional actin filaments, which continuously polymerise and depolymerize, allow the cell to create a pushing force against the substrate via focal adhesion, which are mechanical links between internal actin bundles and the external surface~\cite{sarangi2016coordination}.
In doing so, the cell polarizes its actin filaments and tends to elongate in the direction of motion, 
creating protrusions (lamellipodia) \cite{mitchison1996actin,alberts2002molecular}.
This suggests a minimal physical picture of cellular motility where each cell experiences a net active force
\[
    \vect F_i^{\text{pol.}} = \alpha \vect p_i
\]
in the direction of its polarity $\vect p_i$, where $\alpha$ is a parameter denoting the strength of the focal adhesion (Fig.~\ref{fig:tissue}(a)).
This force contributes to the force-balance equation~\eqref{eq:cells:force_balance} and propels the phase field $\phi_i$ uniformly in the direction of $\vect p_i$.
A simple non-trivial implementation can be obtained by introducing for each cell a new internal degree of freedom $\theta_i^{\text{pol.}}$ describing the direction of its polarity $\vect p_i = (\cos \theta_i^{\text{pol.}}, \sin \theta_i^{\text{pol.}})$ and assume that the polarization angle relaxes diffusively towards an arbitrary vector $\vect d_i$~\cite{peyret2019sustained}.
This leads to the following equation of motion
\begin{equation}
\label{eq:alignment}
\partial_t \theta_i^{\text{pol.}} = - J^{\text{pol.}}\, \abs{\vect d_i}\, \measuredangle(\vect p_i, \vect d_i) + (2 D^{\text{pol.}})^{\frac 1 2} \, \eta_i\,,
\end{equation}
where $\measuredangle(\vect p_i, \vect d_i)$ is the angle between $\vect p_i$ and $\vect d$, $\eta_i$ is Gaussian white noise, and the positive constants $J^{\text{pol.}}$ and $D^{\text{pol.}}$ are the strength of the alignment torque and of the rotational diffusivity, respectively.
This defines a dynamics similar to an Ornstein-Uhlenbeck process and has been well studied in the context of active systems~\cite{szabo2006phase,giavazzi2018flocking}.

There are multiple interesting choices for the definition of the aligning direction $\vect d_i$ and there is as of today no clear microscopic or dynamical basis favouring one of them~\cite{alert2020physical}.
A simple choice that is consistent with the mechanism of contact inhibition of locomotion (CIL)~\cite{smeets2016emergent} is to assume that the polarity aligns to the direction of the total force exerted on a cell's interface, namely $\vect d_i = \vect F^{\text{int.}}_i$.
Such a coupling was shown to describe accurately patterns of sustained oscillations observed experimentally in systems of confined MDCK cells~\cite{peyret2019sustained}.
Other interesting possibilities are to assume that the polarity aligns either to the total velocity $\vect v_i$ of each cell or to its main axis of elongation, and an extensive evaluation of these two choices is presented in~\cite{alert2020physical,Zhang2020}.
At high enough alignment strengths, the former case shows a Vicsek-type phase transition where all cells move in the same direction, while the latter exhibits unjamming to a liquid-like state~\cite{Zhang2020}.
Finally, the purely diffusive dynamics of the system ($J^{\text{pol.}} = 0$) is interesting in its own right and it has been shown that densely packed cells can show “bursts” of motions as they quickly relax from their deformed shape in this case~\cite{palmieri2015multiple}, while confluent monolayers exhibit a solid-liquid phase transition as $D^{\text{pol.}}$ is increased~\cite{loewe2019solid,Zhang2020}.
Further research should investigate potential microscopic bases for such alignment dynamics of the polarization and put on a firmer footing.

\subsection{Active nematic driving}
\begin{figure}[h]
\centering
  \includegraphics[width=1.0\linewidth]{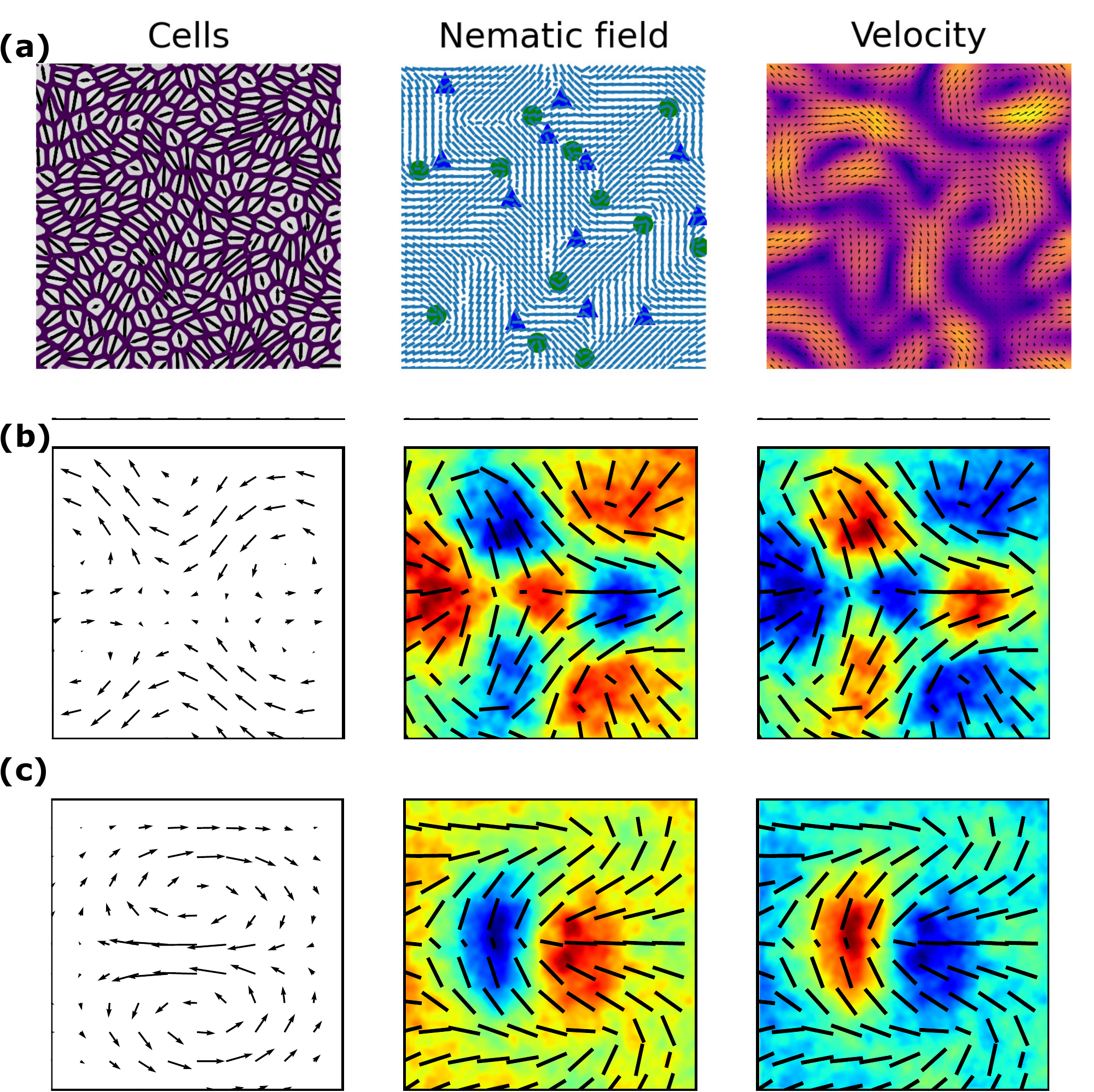}
  \caption{{\bf Emergent features in a phase field model of cell monolayer.} (a) Emergence of active turbulence. Shown here are the cells in confluence, their associated coarse-grained nematic field, and the corresponding velocity field. The nematic field describes the local orientation of the cells defined as their elongation direction. The singularities in the orientation field are marked by green circles for $+1/2$ and blue triangles for $-1/2$ topological defects. The velocity field is colored by the normalized magnitude of velocity ranging from $0$ (purple) to $1$ (yellow). (b, c) Average velocity field, isotropic stress, and pressure around $-1/2$ (b) and $+1/2$ topological defects (c). Figures adapted from~\cite{mueller2019emergence}.}
  \label{fig:tissueres}
\end{figure}
While polar driving has been extensively studied theoretically in the context of single cells, there is strong evidence that descriptions based on active nematic liquid crystals offer a compelling framework to understand the collective behaviour of cell monolayers~\cite{saw2017topological,blanch2018turbulent,mueller2019emergence}.
Nematic driving can be introduced following~\cite{mueller2019emergence} by rewriting the interface force~\eqref{eq:cells:passive} as
\begin{equation} \label{eq:cells:Fsigma}
    \vect F^{\text{inter.}}_i = \int \dd \vect{x}\, \phi_i \vecnabla \matr \Pi^{\text{tissue}} = - \int \dd \vect{x} \, \matr \Pi^{\text{tissue}} \vecnabla \phi_i,
\end{equation}
where $\matr \Pi^{\text{tissue}}$ is a tissue stress tensor that can be written in the usual fashion as
\begin{equation} \label{eq:cells:sigma}
    \matr \Pi^{\text{tissue}} =  - p \id - \zeta \matr Q.
\end{equation}
The pressure $p = \delta (\mathcal F_{\text{rep.}} - \mathcal F_{\text{CH}} - \mathcal F_{\text{area}}) / \delta \phi_i$ can be read directly from equation~\eqref{eq:cells:passive}, while $\matr Q = \sum_i \phi_i \matr Q_i$ is a newly introduced tissue nematic tensor written as a weighted sum over the contributions of the individual cells.
Such a definition has two main advantages: (i) it bridges the gap between local (cell level) and global (tissue level) properties and (ii) allows for naturally introducing the usual active term $- \zeta \matr Q$ found in continuum theories of active liquid crystals~\cite{Marchetti13,doostmohammadi2018active}.
It can be interpreted as creating a dipolar force density distributed along the cells interfaces such that each cell pushes or pulls its neighbours depending on the direction of their contact area with respect to the nematic tensor~(Fig.~\ref{fig:tissue}(a)).

The simplest choice for $\matr Q_i$ is to introduce an internal nematic degree of freedom  for each cell and define its dynamics similarly to the polar case.
In the continuum theory, most of phenomenology of active liquid crystals depends on a balance between the restoring elastic forces and flow alignment of the nematic tensor.
Writing $\matr Q_i = 2 \left(\uvect n_i^\transp \uvect n_i - \id /2\right)$ with $\uvect n_i = (\cos \theta_i^{\text{nem.}}, \sin \theta_i^{\text{nem.}})$, we can mirror these two components in our model by defining the following dynamics of the angle $\theta_i$:
\begin{equation}
\label{eq:theta}
    \partial_t \theta^{\text{nem.}}_{i} = K^{\text{nem.}} \tau_i + J^{\text{nem.}} \omega_i,
\end{equation}
where the torques are given by
\begin{align}
    \label{eq:torques}
    \tau_i = \frac 1 \lambda \int \dd \vect x\;  \phi_i\, \matr Q \wedge \matr Q_i, \quad
    \omega_i = \int \dd \vect x\; \vect v \wedge \nabla \phi_i.
\end{align}
We have defined the tissue velocity as $\vect v = \sum_i \phi_i \vect v_i$ and $\matr A \wedge \matr B = A_{xx} B_{xy} - A_{xy} B_{xx}$ for symmetric traceless matrices $\matr A$ and $\matr B$.
The first torque $\tau_i$ aligns $\matr Q_i$ to the tissue nematic tensor $\matr Q$ and induces an elastic restoring force favouring the homogeneous state.
The second torque $\omega_i$ rotates $\matr Q_i$ with the local vorticity computed as the integral of the neighbouring cells velocity projected on the cell interface.
Toghether with the active term, these torques are able to drive the cells out of equilibrium and reproduce the bend and splay instabilities observed in continuum theories of active nematic liquid crystals (Fig.~\ref{fig:tissue}(b)).
In a confluent epithelium, one can also check that the total force is approximately zero and the system does not develop any total velocity under periodic boundary conditions.
In particular this means the transition to collective movement in is different from the Vicsek-type phase transitions observed with polar driving~\cite{giavazzi2018flocking}.

Another interesting possibility explored in~\cite{mueller2019emergence} is to avoid introducing supplementary degrees of freedom altogether and define an active coupling that is directly proportional to the shape deformations of the individual cells.
This is motivated by recent studies indicating that the local deformation of cells provides a suitable nematic order parameter that allows the description of the dynamics of epithelial cells using theories of active liquid crystals~\cite{doostmohammadi2015,duclos2017topological,saw2017topological,duclos2018spontaneous,blanch2018turbulent}.
Such a connection is quite surprising because individual epithelial cells on a substrate are typically not elongated and have a well-defined direction of movement, suggesting polar rather than nematic driving.
In this setting, the tissue nematic tensor is defined as $\matr Q = \sum_i \phi_i \matr S_i$, where $\matr S_i$ is the deformation tensor of cell $i$ given by
\begin{equation} \label{eq:cells:shape}
    \matr S_i = - \int\dd\vect x\left( (\vecnabla \phi_i)^\transp \vecnabla \phi_i - \frac 1 2 \tr \left[(\vecnabla \phi_i)^\transp \vecnabla \phi_i \right]\right).
\end{equation}
Equation~\eqref{eq:cells:shape} defines a $2\times2$ matrix whose eigenvalues and eigenvectors describe the strength and orientation of the main deformation axes of each cell, see~\cite{mueller2019emergence,asipauskas2003texture} for details.
This corresponds to a nematic tensor with order parameter and director given by the largest eigenvalue and its associated eigenvector.
With this simple definition, the system shows an activity-driven transition to non-zero nematic order and flows for high enough activity strengths $\zeta$.
The spontaneous creation of defects in the coarse grained nematic field as well as observed patterns of flows and mechanical stresses around topological defects are accurately predicted, see Fig.~\ref{fig:tissueres}.
This points to a strong connection between the shape deformation of cells and their active behaviour, irrespective of the corresponding microscopic mechanism.

\section{Summary}
We have presented an introduction to phase field modeling of active systems, providing examples from both continuum biphasic active materials and agent-based models of cellular monolayers. A general framework that allows the introduction of various forms of coupling to the phase field was developed. In particular, it was shown how hydrodynamic effects can be included in phase field models and how tensor order parameters such as active particles orientation field or polymer conformation (describing viscoelastic effects) can be coupled to the equations of motions of the phase field in order to construct complex models of active interfaces. Furthermore, we have described how using multiple phase fields can be used to construct a versatile model of cellular monolayers, where each cell is described as an active deformable droplet using an individual phase field. Applications of this approach to modeling collective cell motion and capturing cell-cell interaction forces were described.

As we gain more knowledge of complex spatiotemporal features of active systems, understanding the dynamics of physical interfaces between different active materials as well as between active matter and its surrounding media becomes more and more important.
Such understanding will be of prime importance to tackle biophysical problems such as bacterial biofilm development, cellular invasion, and morphogenesis.
Building versatile and predictive computational models of active interfaces will be, together with experimental advances, the determining factor in advancing our understanding of such systems, both by helping to explain the mechanisms behind experimental observations and, even more importantly, by providing predictive tools for experimentally inaccessible scenarios.

\section{Acknowledgement}
The notes here are part of the ``Initial Training on Numerical Methods of Active Matter" organised by MSCA-ITN ActiveMatter (Grant: 812780) funded by the EU H2020 programme. AD acknowledges support from the Novo Nordisk Foundation (grant no. NNF18SA0035142), Villum Fonden (grant no. 29476), Danish Council for Independent Research, Natural Sciences (DFF-117155-1001), and funding from the European Union’s Horizon 2020 research and innovation program under the Marie Sklodowska-Curie grant agreement no. 847523 (INTERACTIONS).

\bibliographystyle{apsrev4-1}
\bibliography{references}
\end{document}